\documentclass[11pt]{article}
\usepackage[utf8]{inputenc}
\usepackage[T1]{fontenc}
\usepackage{fixltx2e}
\usepackage{graphicx}
\usepackage{longtable}
\usepackage{float}
\usepackage{wrapfig}
\usepackage{soul}
\usepackage{textcomp}
\usepackage{marvosym}
\usepackage{wasysym}
\usepackage{latexsym}
\usepackage{amssymb}
\usepackage{hyperref}
\usepackage{color}
\usepackage{listings}
\usepackage{fullpage}
\usepackage[american]{babel}
\usepackage[T1]{fontenc}
\usepackage[utf8]{inputenc}
\usepackage[american]{babel}
\usepackage{xspace}
\usepackage{url} 
\usepackage{titling}

\title{}
\date{}
\hypersetup{
  pdfkeywords={},
  pdfsubject={},
  pdfcreator={Emacs Org-mode version 7.8.11}}

\begin{document}

\def\ie{i.e.,\xspace}
\def\eg{e.g.,\xspace}
\newcommand{\sg}{SimGrid\xspace}%
\setlength{\droptitle}{-2cm}
\sloppy
\title{SimGrid: a Sustained Effort for the Versatile Simulation\\ of
Large Scale Distributed Systems}

\author{
Henri Casanova\\
ICS Dept., U. of Hawai`i\\
Manoa, U.S.A
\and
Arnaud Giersch\\
FEMTO-ST, U. of Franche-Comt\'e \\
Belfort, France
\and 
Arnaud Legrand \\ LIG,  CNRS \\ Grenoble, France
\and 
Martin Quinson\\
Loria, U. de Lorraine\\
Nancy, France
\and 
Frédéric Suter\\
IN2P3 Computing Center, CNRS\\
Lyon-Villeurbanne, France
}

\maketitle
~
\vspace{-2cm}

\section{Introduction}
\label{sec-1}

\vspace{-1mm}
Computers have revolutionized science by fully automating the mathematical
computations involved in modeling activities. This move toward
computational science has proved decisive throughout modern science and
engineering.  The computations underlying these scientific advances are
carried out on very large computing facilities the scale of which is ever
increasing.  For instance, current dedicated High Performance Computing
(HPC) systems can comprise hundreds of thousands of cores. Furthermore,
Grid systems are deployed as transnational federations of large-scale
computing facilities that can aggregate millions of heterogeneous computing
elements in deep hierarchies. Such systems are certainly among the most
complex artifacts ever built by mankind.  Then, cloud computing and
virtualization technologies allow to outsource a computing infrastructure
to dedicated data centers whose physical location is transparent to the
users. Finally, among the most complex large-scale platforms are the
decentralized infrastructures enabled by P2P technologies, which federate
widely heterogeneous compute and storage resources located at the edge of
the network.

While used routinely, these platforms remain for the most part poorly
understood.  Simulation is an appealing approach to study large-scale
distributed systems. Simulation consists in predicting the evolution
of a behavior of the system (both the application and the platform)
through numerical and algorithmic models. Simulation experiments are
\emph{fully repeatable and configurable}, making it possible to explore
``what if'' scenarios without an actual platform
deployment. Furthermore, simulation is often less labor intensive,
costly, and/or time consuming than conducting experiments on a real
platform.  Simulation raises its own issues, among which the question
of the induced simulation bias is critical.  Nevertheless, simulation
has proved useful in most scientific and engineering disciplines, and
holds the same promise for the study of large-scale distributed
computing platforms.

Many simulators have been developed over the last
decade for the simulation of almost every kind of large scale
distributed systems.  However, the field is highly partitioned and
each tool is usually limited to a specific domain, and even to a
specific study within that domain.  Moreover the simulators, most of
which do not survive beyond the lifetime of a given funded project,
are rarely made available or even detailed in publications. For
instance, \cite{P2P_survey} points out that out of 141 surveyed papers
that use simulation for the study of P2P systems, 30\% use a custom
simulator, and 50\% do not even report which simulator was used.  The
consequence is that most published simulation results are impossible
to reproduce by researchers other than the authors.  And yet,
simulation results should be easily repeatable by design!

A noticeable exception is the \sg framework \cite{simgrid_web}. This
versatile scientific instrument has been used successfully for
simulation studies in Grid Computing, Cloud Computing, HPC, Volunteer
Computing and P2P Systems. It was shown both more realistic and more
scalable than its major competitors, thus blurring the boundaries
between research domains.  Over the past five years only, this
framework was used by more than 100 distinct authors from 4 continents
in over 60 research articles and Ph.D.  dissertations. Its development
has been sustained for nearly 15 years. In Section
\ref{sec:sg_presentation} we briefly retrace the history and evolution
of \sg.  In Section \ref{sec:sustain} we explain how \sg has achieved
sustained development. Finally, Section \ref{sec:ccl} provides a brief
summary of the lesson learned while developing \sg.

\vspace{-3mm}
\section{\sg: A Versatile Scientific Instrument}
\label{sec-2}

\label{sec:sg_presentation}
\vspace{-1mm}
\sg is a 15-year old open source project whose domain of application
has kept growing since its inception.  It was initiated in 1998
by H. Casanova at the University of California, San Diego as a way to
factor and consolidate different simulators developed by different graduate students for
studying scheduling algorithms for heterogeneous platforms. The first
version of \sg \cite{simgrid_ccgrid01} made it easy to prototype
scheduling heuristics and to test them on a variety of abstract
applications (expressed as task graphs) and platforms.

In 2003, the work of A. Legrand at the École Normale Supérieure de
Lyon led to the second major release of \sg \cite{simgrid_ccgrid03}
that extended the capabilities of its predecessor in two major
ways. First, the accuracy of the simulation models was improved by
transitioning the network model from a wormhole model to an analytical
fluid model.  Second, an API, called MSG, was added to simulate
generic Communicating Sequential Processes (CSP) scenarios.

The third major release of \sg was distributed in 2005, but major new
features appeared in version 3.3 in April 2009 \cite{simgrid_uksim08},
after contributions from M. Quinson and F. Suter. These features
include a complete rewrite of the simulation core for better
modularity, speed, and scalability, the possibility to attach traces to
resources to simulate time-dependent performance characteristics as
well as failure events.  Two new user interface were added: SimDag,
which revives some of the original concepts from the first release, and GRAS,
which allows users to develop/test code in simulation and then
run it unmodified on a real platform.

From early 2009 up until today, and thanks to securing large amounts of funding, the
domain of application of \sg has been extended to P2P, HPC and Cloud
infrastructures and applications. New models and concepts have been 
added to the code base and a new API for the simulation of unmodified
MPI applications was proposed. In the same period, the internals
became more mature thanks to the work of A. Giersch and the heavy use
of a multi-platform/multi-OS testing infrastructure. Finally the
simulation framework itself now comes with an ecosystem of tools
ranging from the generation of simulation scenarios to the
visualization of simulation results.

\begin{figure}[hbt]
\centering
\includegraphics[width=.8\linewidth]{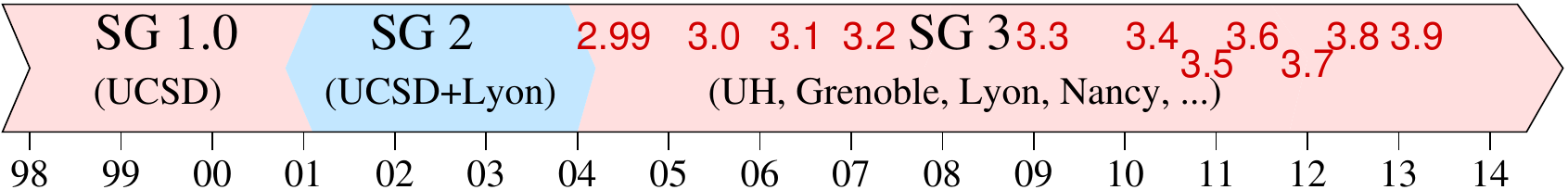}%

\vspace{-1.5mm}\caption{The \sg development timeline.}
\label{fig:dev_timeline}
\end{figure}

Fig. \ref{fig:dev_timeline} summarizes the release timeline of \sg
since its very first version and indicates the main locations of the
members of the development team.  In 15 years, \sg evolved from a
project developed in-house at a single laboratory into an
international collaboration that involves more than 20 active
contributors. It also evolved from a very domain-specific simulator
into a \emph{versatile scientific instrument}, whose performance and
accuracy are continuously and thoroughly assessed, for the study of
large scale distributed systems \cite{simgrid3}.

\vspace{-3mm}
\section{Sustaining  \sg as a Research Software \label{sec:sustain}}
\label{sec-3}

\vspace{-1mm} The sustained evolution of \sg was made possible by
different factors: obtaining \emph{recurring funding}, which provides
manpower, \emph{improving the code base}, which can be accomplished via
both custom and standard tools, and retaining and growing the \emph{user base}.

\vspace{-3mm}\paragraph*{Securing Recurrent Funding and Support --} As
mentioned earlier, most simulators do not survive beyond the lifetime of
the grant that funded its development. However, the \sg project was
developed without specific financial support for about eight years. The
development effort was then supported by A. Legrand and M. Quinson alone.
They used \sg early on for their Ph.D. research, and then continued to use
and develop it as part of their subsequent research activities.  But using
and developing a tool both for one's own research and for reaching out to a
larger user community requires additional manpower.  Since 2006, the \sg
project was funded by four major grants. We coupled funding requests on
scientific aspects, which allowed us to extend the domain of application
\sg and enlarge the community of research that use it and contributed to
it, with funding requests targeted to the technical support and development
necessary to strengthen the code base.  The USS
SimGrid\footnote{\href{http://uss-simgrid.gforge.inria.fr}{http://uss-simgrid.gforge.inria.fr}
 } project (2009-2012)
aimed at simulating P2P and volunteer computing systems, while the
SONGS\footnote{\href{http://infra-songs.gforge.inria.fr}{http://infra-songs.gforge.inria.fr}
 } project (2013-2015) add
two more application domains: HPC and Cloud computing.  This funding
provided manpower, i.e., Ph.D students, postdocs, and engineers to tackle
the challenging issues raised by the new application domains. Two more
grants, solely dedicated to software engineering, were also secured over
that time period. They led to improved implementation, portability, testing
capabilities, packaging, and documentation.

The key to sustainability, at least in our case, was to request funding
from different sources, i.e., national funding agency or academic
institutes and universities, both for scientific and for technical aspects
of the project. The successes achieved in the scope of a given grant
generally helps to secure a subsequent one.  However, the ability and
willingness to sustain the research and software development efforts in the
(temporary) absence of funding have also been essential.

\vspace{-3mm}\paragraph*{Increasing Code Stability and Quality --}
Going from an in-house project at a single laboratory to a widely distributed and used Open Source simulation
software requires to make the code base as robust, stable,
available, usable, and trustworthy, as possible. While the research
activities improve the models and simulations quality and make the
obtained results trustworthy, a large body of engineering work has
been done about the other requirements. Since March 2005, the project
is hosted on a forge \cite{simgrid_web} operated by Inria, the French
Institute for Computer Science, which ensures a sustained
maintenance. It offers classical services, such as a git
repository, bug tracking systems, mailing lists, or web hosting. 
\sg is distributed along with more than 250 unit tests and examples
that are managed by a custom tool called \emph{tesh} (testing environment
shell). Tesh runs the complete test suite and checks for discrepancies
in output that indicate the apparition of bugs in the code. Thanks to
dashboard\footnote{\href{http://cdash.inria.fr/CDash/index.php?project=SimGrid}{http://cdash.inria.fr/CDash/index.php?project=SimGrid}
 }
and continuous integration\footnote{\href{http://ci.inria.fr}{http://ci.inria.fr}
 } platforms, provided by Inria, the code base is regularly
tested on several architectures and operating systems
combinations. Code coverage and memory leaks are also monitored thanks
to these platforms. All these tools helped to increase the release
frequency seen in Fig. \ref{fig:dev_timeline}.

The key point is that while research usually is the main focus of the
developers of a scientific software, sustainability requires an
important amount of engineering work that cannot be
neglected. Leveraging available tools and platforms that can ease this
task is crucial.

\vspace{-3mm}
\paragraph*{Fostering a User Base --}
The main reason for a scientific software to perdure is its user community,
which should be growing and encouraging user-driven contributions.
The most standard way to interact with users is via a mailing list. More than
a hundred people subscribe to
\texttt{simgrid-user@lists.gforge.inria.fr} and \emph{every} message is
answered in a reasonable delay. This is a good way to retain existing
users and newcomers, but getting more users requires more proactive
communication actions. Over the last five years we have developed a
comprehensive set of tutorials, which are available online, and have greatly
improved the user documentation. \sg is also presented on the Inria booth at
the SuperComputing conference since 2009. We also initiated a
series of events, the \sg Users Days, that are an opportunity for
advanced users to meet developers and discuss the latest features of
SimGrid and currently explored research areas as well as to influence
future developments. The last edition in June 2013 was a bit
different. Users and developers of \sg participated to a 3-day ``coding sprint.''
This was an opportunity to solve existing issues and to address new
challenges jointly with users on very concrete use cases.

We believe that communicating about the tool outside of regular
scientific publications and making efforts to interact with users on a
regular basis are key actions to make \sg perdure and evolve beyond the
limits of its original developers' domain of research.

\vspace{-3mm}
\section{Conclusion  \label{sec:ccl}}
\label{sec-4}

\vspace{-1mm}
In this paper we presented \sg, a toolkit for the versatile simulation
of large scale distributed systems, whose development effort has been
sustained for the last fifteen years. Over this time period \sg has
evolved from a one-laboratory project in the U.S. into a scientific instrument
developed by an international collaboration. The keys to making this evolution possible
have been \emph{securing of funding}, \emph{improving the quality of the software}, and \emph{increasing the user base}. In this paper we have described how we have been able to make advances on all
three fronts, on which we plan to intensify our efforts over the upcoming years.

\def\raggedright{}
\vspace{-3mm}
\small

\bibliographystyle{abbrv}
\bibliography{simgrid_wssspe13}

\end{document}